\documentclass{llncs}
\usepackage[utf8]{inputenc}
\usepackage{xcolor}
\usepackage{graphicx}
\usepackage{subfigure}
\usepackage{soul}

\title{Relative compression of trajectories}

\author{Nieves R. Brisaboa\inst{1} \and Travis Gagie\inst{2} \and Adri\'an G\'omez-Brand\'on\inst{1} \and Gonzalo Navarro\inst{3} \and Jos\'e R. Param\'a\inst{1}
}
\institute{
 Universidade da Coru\~na, Computer Science Deparment, Spain. \\
\email{\{brisaboa,adrian.gbrandon,jose.parama\}@udc.es}
\and
School of Informatics and Telecommunications, Diego Portales University, Santiago, Chile.
\email{travis.gagie@mail.udp.cl}
\and
Dept. of Computer Science, University of Chile, Chile. 
\email{gnavarro@dcc.uchile.cl} }

\begin{document}

\maketitle
\begin{abstract}
We present RCT, a new compact data structure to represent trajectories of objects. It is based on  a relative compression technique called Relative Lempel-Ziv (RLZ), which compresses sequences by applying an LZ77 encoding with respect to an artificial reference. Combined with $O(z)$-sized data structures on the sequence of phrases  that allows to solve trajectory and spatio-temporal queries efficiently. We plan that RCT improves in compression and time performance the previous compressed representations in the state of the art.

\end{abstract}

\section{Introduction}

% Para que se va a usar

Relative compression techniques were designed to exploit the redundancy of highly repetitive datasets. Those techniques represent a large set of sequences with respect to another sequence called reference. Since the differences between the reference and the rest of elements are small, relative compression saves much space in highly repetitive datasets. These techniques were long applied over DNA collections, however it can be extended to other kind of highly repetitive datasets. For example, in datasets of moving objects over networks (taxis) or with predefined routes between different points (planes or boats), objects follow the best-known route between the origin and the final position, therefore the trajectories tend to be similar and repetitive. The aim of this study is to exploit the repetitiveness of object trajectories and support efficiently spatio-temporal queries by using relative compression.

% Un pequeño resumen de nuestra estructura

The best-known relative compression technique is Relative Lempel-Ziv (RLZ) \cite{kuruppu2010relative}, which compresses each sequence of a dataset by applying an LZ77 \cite{ziv1978compression} parsing with respect to a reference. The reference can be a representative sequence of the dataset or an artificial reference built by parts from other sequences. RLZ gets a good compression ratio and supports efficient random access to the original sequence. Therefore, we can compress trajectories with RLZ, it allows us to retrieve the trajectory of an object at a given interval of time, however RLZ cannot solve efficiently spatio-temporal queries, for example, it cannot retrieve the objects within a region during an interval of time.

In this work, we propose a structure based on RLZ which compresses the set of sequences in a dataset with respect to an artificial reference built using the technique proposed in \cite{liao2016effective}. After the construction of the reference, for each trajectory, the LZ77 parser generates \textit{z} phrases with respect to the reference. By combining existing data structures for trajectories \cite{BGGBNPspire17} built on the
reference with $O(z)$-sized data structures on the sequences of phrases, we
offer the same functionality of previous work \cite{BGGBNPspire17,brisaboa2016gract} within RLZ-bounded space. We plan that this arrangement will obtain the best from
both previous work: the low space requirement of GraCT \cite{brisaboa2016gract} and the speed of ContaCT \cite{BGGBNPspire17}.

\section{Background}

% Explicar RLZ
\subsection{Relative Lempel-Ziv}
Let $s = [1..n]$ be a sequence of length $n$ called \textit{source} and $r = [1..m]$ be a sequence of length $m$ called \textit{reference}, where $m \leq n$. The Relative Lempel-Ziv (RLZ) compresses $s$ by using an LZ77 parsing with respect to $r$. As a consequence of the parsing step, we obtain \textit{z} phrases which represent $s$. Therefore $s$ can be represented with $z$ phrases $w_1w_2w_3...w_z$, and every phrase is stored as a pair $(p_i, l_i)$ where $p_i$ is the starting position of the $i$-th phrase at $r$ and $l_i$ its length.

For example, with $s = tgacacacttg$ and $r = tggcacttgat$, RLZ represents $s$ with three phrases. The first phrase is $w_1=tga$ and it is represented with the pair $(8,3)$ because it appears at position $8$ in $r$ and $|w_1|=3$. $w_2=cac$ beginning at position at position $4$ and with length $3$, hence it is encoded with $(4,3)$. Finally, we obtain the pair $(5,5)$ which corresponds with $w_3=acttg$.

\subsection{GraCT}
% Explicar GraCT

GraCT \cite{brisaboa2016gract} is a compact data structure designed to store trajectories and support spatio-temporal queries. It assumes regular timestamps and stores trajectories using two components. At regular periods of time, it represents the position of all the objects in a structure called \textit{snapshot}. The positions of objects between snapshots are represented in a structure called \textit{log}.

Let us denote $Sp_{k}$ the snapshot representing the position of all the objects at timestamp $k$. Given a parameter $d$, which specifies the distance between two consecutive snapshots, $Sp_{k}$ and $Sp_{k+d}$, there is a log for each object, which is denoted ${\cal L}_{k,k+d}(id)$, being $id$ the identifier of the object. 
 The log stores the differences of positions compressed with {\em RePair}  \cite{larsson2000off}, a grammar-based compressor. In order to speed up the queries over the resulting sequence, the nonterminals are enriched with additional information, mainly the MBB of the trajectory segment encoded by the nonterminal. 

 Each snapshot is a binary matrix where a cell set to 1 indicates that one or more objects are placed in that position of the space. To store such a (generally sparse) matrix, it uses a $k^2$-tree \cite{ktree}. The $k^2$-tree is a space- and time- efficient version of a region quadtree \cite{Sam2006}, and is used to filter the objects that may be relevant for queries which involve spatial areas.

\subsection{ContaCT}
% Explicar ContaCT

ContaCT is based on GraCT, hence it keeps the same components: snapshots and logs. The main differences  are in the log. Instead of a log compressed with {\em RePair}, the differences of positions are stored continuously using two bitmap  per axis, one for positive movements and another for negative movements.  For the $x$-axis, we have the bitmap $x_p$ where we store how many positions an object moves to the right in each timestamp and the bitmap $x_n$ which stores the movements to the left. Two additional bitmaps are required to store the movements up and downwards.   In those bitmaps, the movements are encoded in unary. For example, if an object moves two positions to the right, its representation is $x_p=001$ and $x_n=1$ because it moves $0_{10}=1_1$ positions to the left.

In order to compute the position of an object in GraCT, we have to perform a sequential traversal of the log from the closest snapshot up to the queried timestamp. ContaCT avoids this traversal, we can compute the position of an object by using \textit{select} \cite{j-ssds-89} operations over those bitmaps, which can be solved in constant time using an extra space of $o(n)$ bits.

%Añadir comparacion ContaCT vs GraCT

\section{Relative compression of trajectories (RCT)}

RCT uses snapshots and logs, just like ContaCT and GraCT. As in the previous structures, in RCT the main differences involve the log, which is composed by two parts: an artificial reference and the log of trajectories. 

By using the technique presented in \cite{liao2016effective}, we built an artificial reference which represents well all the trajectories stored in the dataset. After the construction of the artificial reference, the trajectories are compressed with RLZ. In order to speed up the queries, an structure similar to ContaCT is built on the artificial reference and $O(z)$-sized structures are added on the sequences of phrases, being $z$ the number of phrases of LZ77 parsing.

\subsection{Artificial reference}

The phrases obtained after applying RLZ compression are pointers to the artificial reference, therefore most of the queries have to be solved on the artificial reference. For this reason, we need a mechanism to compute efficiently two queries:
\begin{itemize}
    \item \textit{movement(i,j)}, computes the movement performed in the reference from the time instant \textit{i} to the time instant \textit{j}. It returns the pair $\{\Delta_x(i,j),\Delta_y(i,j)\}$ where $\Delta_x(i,j)$ is the difference in $x$-axis from $i$ to $j$, and $\Delta_y(i,j)$ the equivalent in $y$-axis .
    \item \textit{MBB(i,j)}, computes the minimum bounding box of the movements represented by the reference between time instants $i$ and $j$. The value returned, $\{x_{min}, y_{min}, x_{max}, y_{max}\}$, is relative to the position of the object at time instant $i-1$.
\end{itemize}

\begin{figure}[t]
	\centering     %%% not \center
\subfigure[\texttt{Artificial reference}]
{\label{fig:refer}\includegraphics[width=0.45\textwidth]{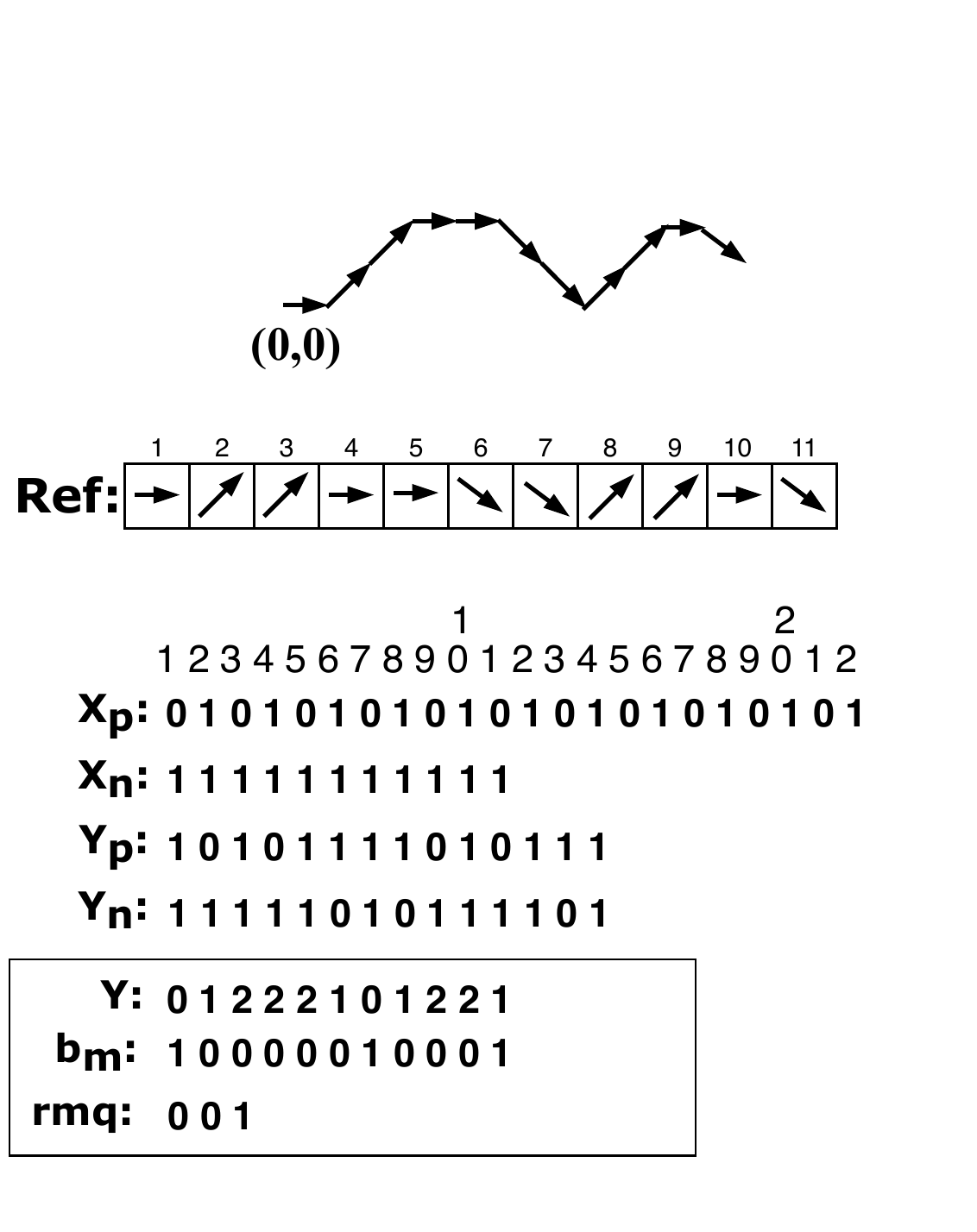}}
\subfigure[\texttt{Log of trajectory}]
{\label{fig:traj}\includegraphics[width=0.45\textwidth]{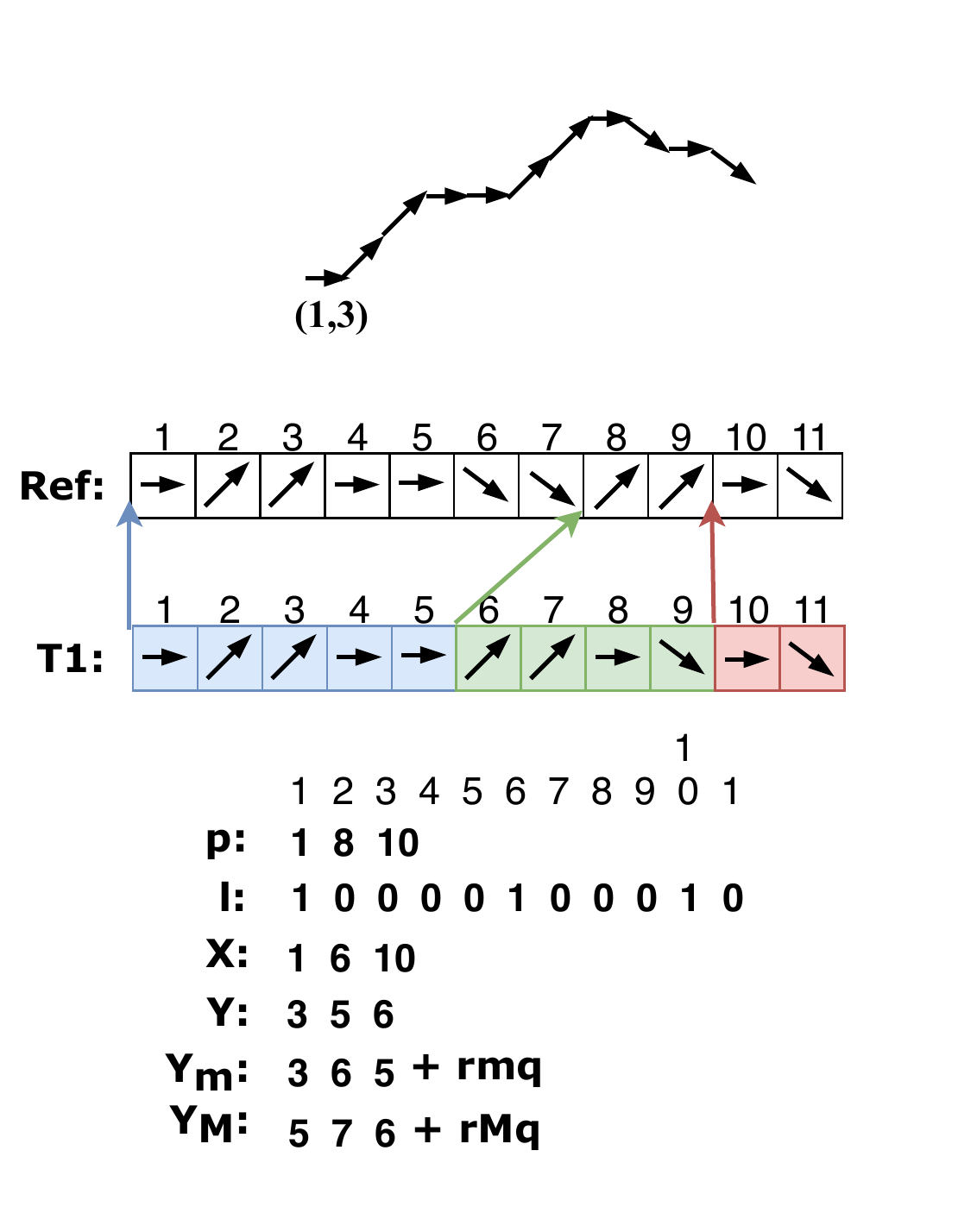}}
\caption{Example of log.}
\label{fig:log}
\end{figure}

In order to support \textit{movement(i,j)}, we add the bitmaps of ContaCT. Recall that all the differences are represented in unary, as we can observe in Figure \ref{fig:refer}. This representation allows that the number of zeroes before the $i$-th 1 corresponds with the number of movements (upwards, downwards, to the right, or to the left) depending on the bitmap. from the initial time instant to the $i$-th movement, denoted as $b.\Delta(i)$,  being $b$: $x_p, x_n, y_p$, or $y_n$. That difference can be computed as $select_1(b, i) - i$, which is solved in constant time using an extra space of $o(n)$ bits. Hence, the number of movements in the bitmap $b$ from $i$ to $j$ is computed as $b.\Delta(i,j) = b.\Delta(j) - b.\Delta(i)$. Therefore, $\Delta_x(i,j)= x_p.\Delta(i,j) - x_n.\Delta(i,j)$ and $\Delta_y(i,j) = y_p.\Delta(i,j) - y_n.\Delta(i,j)$.

To solve \textit{MBB(i,j)}, we need to compute the maximum and minimum in each axis. A naive approach could be store a range minimum query structure \textit{rmq} and a range maximum query structure \textit{rMq} per axis. Both structures only return the index of the minimum/maximum, hence they do not store the values, and each one takes an space of $2n$ bits. For example, the minimum value could be computed as $\Delta_{axis}(i-1, rmq(i,j))$. 
However, $n$ is the size of the trajectory, which can be very large. 

Taking into account that most of the time the objects move in a constant direction, we can sample the local minimums/maximums per axis and mark in a bitmap the movements where the local minimums/maximums appears. For example, in Figure \ref{fig:refer} we can compute the $y_{min}$ of \textit{MBB(5,11)}, we have the bitmap $b_m$ which locates the local minimums and the $rmq$ which returns the index of the local minimum.

First, we compute where we have to run the $rmq$ operation from $rank_1(b_m, i-1)+1=rank_1(b_m, 4)+1=2$ to $rank_1(b_m,j) = rank_1(b_m, 11) = 3$. $rmq(2,3)=2$ and it corresponds with the movement $select_1(b_m, 2) = 7$. Finally, as it is a minimum local, we have to compare the extreme values $Y[5]$ and $Y[11]$ against $Y[7]$ and return the minimum of them. We do not store $Y$, but we can compute $Y[k]=\Delta_y(i-1,k)$ in constant time. The $\Delta_y$ is computed with $i-1$ to obtain a 
relative value. We repeat this step to compute $x_{min}$, $x_{max}$ and $y_{max}$.

\subsection{Log of trajectories}

The trajectories are compressed with RLZ, as we can observe in Figure \ref{fig:traj}, therefore each trajectory $T_i$ is represented with $z$ phrases: $w_1w_2w_3...w_z$. Recall that each phrase is a pair $(p_i, l_i)$ where $p_i$ is the starting position of the $i$-th phrase in the reference and $l_i$ its length. We store the information of the pairs separately, $p_i$ values in an array $p$ and we mark in the bitmap $l$ the beginning of all the $z$ phrases at $T_i$. We store for each $w_i$ the previous position $(x_i,y_i)$. In addition, we store the time instant $t_s$ when the trajectory starts. With this information we can compute the position of an object at $t_q$ in constant time, as we show in the next section. 

The minimum value in $y$-axis of each $w_i$ is stored by the array $Y_m$ with an $rmq$ structure, and the same with the maximum value in an array $Y_M$ with an $rMq$ structure. This structures are replicated per axis, hence we can compute the $MBB$ which involves each $w_i$. As we will explain later, it speeds up the time-interval queries.

\section{Queries}

\subsection{Search object}

Given at time instant $t_q$ and an object identifier $id$ we access the log ${\cal L}(id)$ to retrieve the position at $t_q$. First we perform a $j = rank_1(l, t_q - t_s)$, thus we know $w_j$ contains the result, and it is stored inside $w_j$ at position $k=t_q-t_s-select_1(l,j)$. By accessing the reference we compute the position as $(x_j, y_j) + movement(p[j]-1, p[j]+k)$.

\subsection{Trajectory}
The operation trajectory returns the position of a given object between two time instants: $t_{start}$ and $t_{end}$. It can be solved by computing the position with \textit{search object} at $t_{start}$ and applying \textit{movement(i,i+1)} for every $i \in w_j$, where $w_j$ belongs to the set of phrases which contains the time instants between $t_{start}$ and $t_{end}$.

\subsection{Time slice}
Let us define a region $R=[x_1, x_2]\times[y_1, y_2]$. Time slice returns the objects within $R$ at a given time instant $t_q$. In order to solve this query we have to consider the maximum speed of the dataset, $speed_{max}$. The algorithm starts by obtaining the candidates from the previous snapshot ${\cal S}_p$ at $t_p$, it means, all the objects which are contained in $R'$, where $R'$ extends $R$ in all directions $speed_{max} \times (t_q - t_p)$. Finally, we compute the position at $t_q$ of every candidate using the operation \textit{search object} and return those candidates that are contained in $R$.

\subsection{Time interval}
\begin{figure}[t]
	\centering     %%% not \center
	\includegraphics[width=0.45\textwidth]{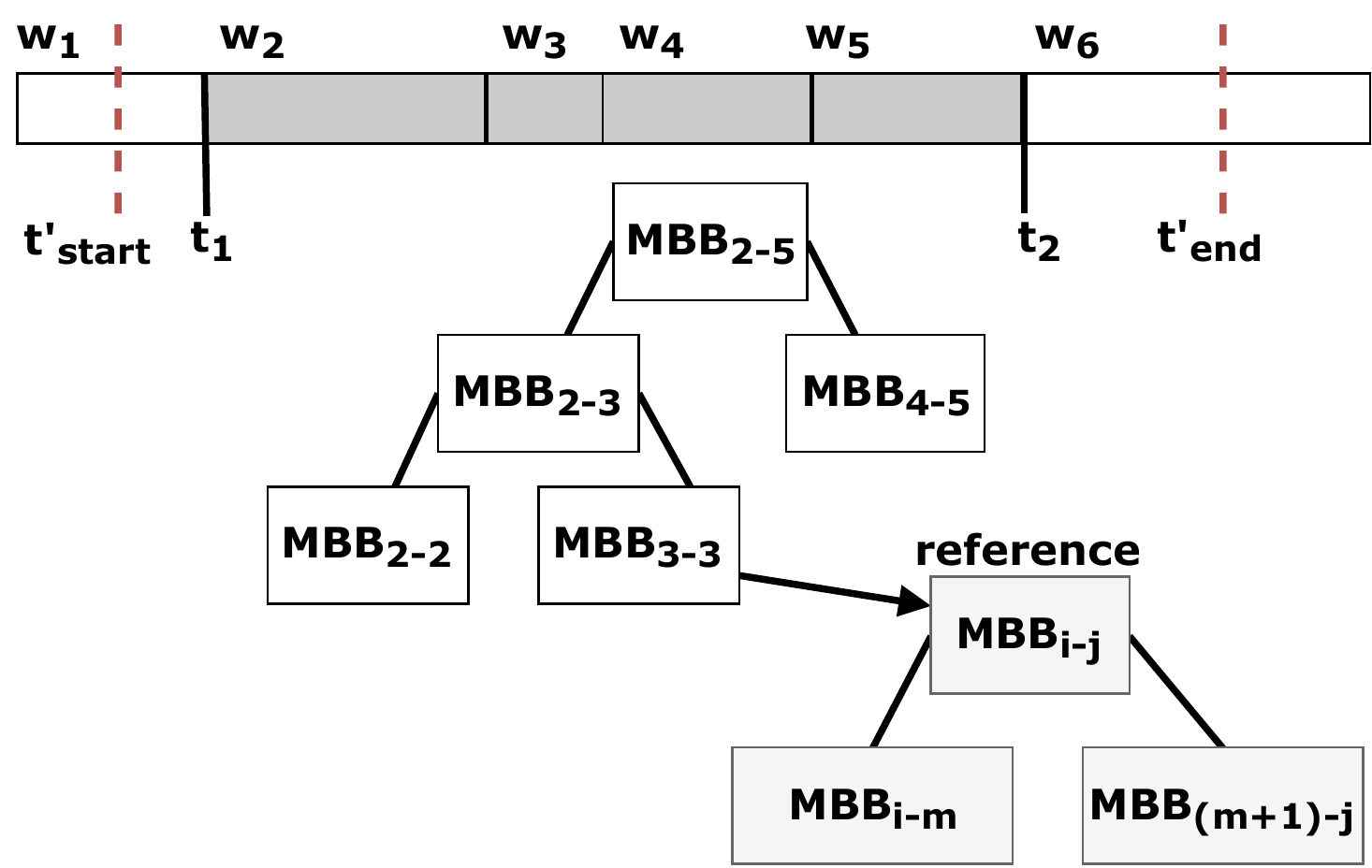}
\caption{Time interval query.}
\label{fig:interval}
\end{figure}

The time interval query returns those objects which are in a given region $R$ at any time instant $t_i \in [t_{start}, t_{end}]$. To solve this query we split the interval into as many intervals $t_{start}', t_{end}'$ as portions of the log between two snapshots overlaps $[t_{start}, t_{end}]$ . Then, this portions can be solved similar to time-slice.  Firstly, we obtain the candidates from the previous snapshot ${\cal S}_p$ at $t_p$ by extending $R$ in all directions $speed_{max} \times (t_{end}' - t_p)$. For each candidate, we check the phrases which overlap the interval $t_{start}', t_{end}'$. As we can observe in Figure \ref{fig:interval}, some phrase are completely included in $[t_{start}', t_{end}']$ but others can be partially included. In the worst case, there are one interval $[t_1, t_2]$ where the phrases are completely covered and two extreme partially covered intervals ($[t_{start}', t_1]$ and $[t_2, t_{end}']$).

Since we are storing the minimum and maximum of each phrase per axis, we can compute the $MBB$ covered by $[t_1, t_2]$ in constant time. First, we need to know the interval of phrases equivalent to $[t_1, t_2]$, that interval is between $w_{s} = rank_1(l, t_1-1)+1$ and $w_{e}= rank_1(l, t_2)$. By computing  $X_m.rmq(w_{s}, w_{e})$ and $X_M.rMq(w_{s}, w_{e})$ we obtain the minimum and maximum of the $MBB$ for the $x$-axis, respectively. Analogously, we obtain the minimum and maximum of $y$-axis. Then, we check if the $MBB$ is contained in $R$, in that case we add the object to the solution and stop the search. If the $MBB$ intersects with $R$, we run this process recursively splitting the interval $[w_s, w_e]$ into two halves. On the other hand, if $MBB$ is outside $R$ and they do not intersect, we stop processing the actual interval and we process the next one.

%\textcolor{red}{qué pasa cos extremos???? de $t'_{start}$ a $t_1$ e de $t_2$ a $t'_{end}$??}

Once, we have to process only one phrase, we run the binary search in the reference. We split it in two halves and we continue recursively computing the $MBB$ as $MBB(i,j) + (x_p, y_p)$, where $MBB(i,j)$ is the relative $MBB$ and $(x_p,y_p)$ the previous location of the object. When we have not found any $MBB$ completely contained in $R$ between $[t_1, t_2]$, we repeat these steps on the reference for the partially covered intervals.

% Time Interval (largo)
\bibliographystyle{splncs03}
\bibliography{biblio}

\begin{thebibliography}{1}
\providecommand{\url}[1]{\texttt{#1}}
\providecommand{\urlprefix}{URL }

\bibitem{BGGBNPspire17}
Brisaboa, N., Gagie, T., G{\'o}mez-Brand{\'o}n, A., Navarro, G., Parama, J.:
  Efficient compression and indexing of trajectories. In: Proc. 24th
  International Symposium on String Processing and Information Retrieval
  (SPIRE). pp. 103--115. LNCS 10508 (2017)

\bibitem{ktree}
Brisaboa, N.R., Ladra, S., Navarro, G.: Compact representation of web graphs
  with extended functionality. Information Systems  39(1),  152--174 (2014)

\bibitem{brisaboa2016gract}
Brisaboa, N.R., G{\'o}mez-Brand{\'o}n, A., Navarro, G., Param{\'a}, J.R.:
  Gract: a grammar based compressed representation of trajectories. In:
  International Symposium on String Processing and Information Retrieval. pp.
  218--230. Springer (2016)

\bibitem{j-ssds-89}
Jacobson, G.: {Succinct static data structures}. Ph.D. thesis, Carnegie-Mellon
  (1988)

\bibitem{kuruppu2010relative}
Kuruppu, S., Puglisi, S.J., Zobel, J.: Relative lempel-ziv compression of
  genomes for large-scale storage and retrieval. In: International Symposium on
  String Processing and Information Retrieval. pp. 201--206. Springer (2010)

\bibitem{larsson2000off}
Larsson, N.J., Moffat, A.: Off-line dictionary-based compression. Proceedings
  of the IEEE  88(11),  1722--1732 (2000)

\bibitem{liao2016effective}
Liao, K., Petri, M., Moffat, A., Wirth, A.: Effective construction of relative
  lempel-ziv dictionaries. In: Proceedings of the 25th International Conference
  on World Wide Web. pp. 807--816. International World Wide Web Conferences
  Steering Committee (2016)

\bibitem{Sam2006}
Samet, H.: {Foundations of Multimensional and Metric Data Structures}. Morgan
  Kaufmann (2006)

\bibitem{ziv1978compression}
Ziv, J., Lempel, A.: Compression of individual sequences via variable-rate
  coding. IEEE transactions on Information Theory  24(5),  530--536 (1978)

\end{thebibliography}

\end{document}